\documentclass[10pt, conference, letterpaper]{IEEEtran}
\IEEEoverridecommandlockouts
\usepackage{comment}
\usepackage{booktabs}
\usepackage[dvipsnames]{xcolor}
\usepackage{amsmath}
\usepackage{multicol}
\usepackage{graphicx}
\usepackage{acronym}
\usepackage[capitalise,noabbrev]{cleveref}
\usepackage[ruled,vlined,linesnumbered,noresetcount]{algorithm2e}
\usepackage{float}
\usepackage{cite}
\usepackage{amssymb}
\usepackage{url} 

\usepackage{caption}
\usepackage{subcaption}

\usepackage{url} \usepackage[oldenum,olditem]{paralist}

\acrodef{sigla}[eDAFL]{enhanced Dynamic Adaptive FL}
\acrodef{bs}[BS]{Base Station}
\acrodef{los}[LoS]{Line-of-Sight}
\acrodef{nlos}[NLoS]{Non Line-of-Sight}
\acrodef{rss}[RSS]{Received Signal Strength}
\acrodef{qos}[QoS]{Quality of Service}
\acrodef{mmwave}[mmWave]{millimeter Wave}
\acrodef{v2x}[V2X]{Vehicle-to-Everything}
\acrodef{ml}[ML]{Machine Learning}
\acrodef{fl}[FL]{Federated Learning}
\acrodef{dl}[DL]{Deep Learning}
\acrodef{iid}[IID]{Independant and Identically Distributed}
\acrodef{ifca}[IFCA]{Iterative Federated Clustering Algorithm}
\acrodef{pfl}[PFL]{Partial Federated Learning}
\acrodef{cka}[CKA]{Centered Kernel Alignment}
\acrodef{hc}[HC]{Hierarchical Clustering}
\acrodef{cnn}[CNN]{Convolutional Neural Network}
\acrodef{oran}[O-RAN]{Open Radio Access Network}
\acrodef{nn}[NN]{Neural Network}
\acrodef{HSIC}[HSIC]{Hilbert-Schmidt Independence Criterion}
\acrodef{RF}{Radio Frequency}

\def\BibTeX{{\rm B\kern-.05em{\sc i\kern-.025em b}\kern-.08em
T\kern-.1667em\lower.7ex\hbox{E}\kern-.125emX}}

\newcommand{\oldtext}[1]{}

\newcommand{\ie}{\textit{i.e.}}
\newcommand{\etal}{\textit{et al.}}

\author{\IEEEauthorblockN{Lucas Pacheco$^1$$^2$, Torsten Braun$^2$, Kaushik Chowdhury$^3$, Denis Rosário$^1$,  Batool Salehi$^3$, Eduardo Cerqueira$^1$}\\
\IEEEauthorblockA{
$^1$Institute of Technology, Federal University of Pará, Brazil\\
$^2$Institute of Computer Science, University of Bern, Switzerland\\
$^3$Department of Electrical and Computer Engineering
Northeastern University, Boston, MA, USA
}\\
Email: \{lucas.pacheco, torsten.braun\}@unibe.ch,
\{denis, cerqueira\}@ufpa.br, 
\{bsalehihikouei, krc\}@ece.neu.edu
}

\title{Dynamic Adaptive Federated Learning for mmWave Sector Selection}

\begin{document}

\IEEEtitleabstractindextext{
\begin{abstract}
Beamforming techniques use massive antenna arrays to formulate narrow Line-of-Sight signal sectors to address the increased signal attenuation in \ac{mmwave}. However, traditional sector selection schemes involve extensive searches for the highest signal strength sector, introducing extra latency and communication overhead.
This paper introduces a dynamic layer-wise and clustering-based federated learning (FL) algorithm for beam sector selection in autonomous vehicle networks called \ac{sigla}. The algorithm detects and selects the most important layers of a machine learning model for aggregation in FL process, significantly reducing network overhead and failure risks. \ac{sigla} also consider an intra-cluster and inter-cluster approach to reduce overfitting and  increase the abstraction level.
We evaluate \ac{sigla} on a real-world multi-modal dataset, demonstrating improved model accuracy by approximately 6.76\% compared to existing methods, while reducing inference time by 84.04\% and model size up to 52.20\%.
\end{abstract}

}

\maketitle

\IEEEdisplaynontitleabstractindextext

\IEEEpeerreviewmaketitle
\acresetall
\section{Introduction}
Beamforming techniques use massive antenna arrays to formulate narrow \ac{los} signal beams in \ac{mmwave} communication, where vehicles must connect to one available beam sector of a given \ac{bs}.
However, current sector selection protocols rely on extensive search procedures to find the sector with the highest signal strength to establish \ac{los} links with low latency, where the search process requires approximately 20~ms for a group of 34 beams  \cite{salehi_flash_2022}.
Hence, the sector selection latency becomes a limiting factor due to high vehicle mobility, obstacles, and inherent fluctuations in the \ac{rss} of sectors, resulting in \ac{nlos} conditions \cite{Mollah_Wang_Fang_2024}.
Traditional centralized learning approaches are inadequate due to privacy constraints with sensitive vehicular sensor data, as well as high mobility causing rapid topology changes, and 3) Network congestion from continuous raw data transmission~\cite{caldarola2022improving,zhu2021delayed}.

\ac{fl} has emerged as a promising paradigm for distributed machine learning in wireless networks, mainly due to its ability to preserve data privacy, reduce communication overhead, and adapt to dynamic environments.
Previous works on \ac{fl} for \ac{mmwave} sector selection in autonomous vehicle environments have shown their efficiency in providing proactive fine-tuning beam alignment to match user locations while enhancing the \ac{qos} \cite{Salehi_Roy_Gu_Dick_Chowdhury_2024}.
However, models are transmitted and aggregated through control channels with wireless connectivity \cite{Lee_Zhang_Avestimehr_2023}.

In this context, layer-wise strategies improve the compression of \ac{ml} by aggregating only selected layers to reduce data transmission while avoiding pruning issues significantly \cite{Karimi_Li_Li_2022, Liu_Elkerdawy_Ray_Elhoushi_2021}.
By grouping vehicles based on the similarity of their data distributions, each cluster can train specialized models to improve convergence and accuracy, as well as handle non-\ac{iid} data distributions  \cite{Ghosh_Chung_Yin_Ramchandran_2020}.
Hence, it is important to predict and select the best \ac{mmwave} sector for a vehicle to connect with higher communication efficiency and reliability while achieving low latency in order to address both the complexity of \ac{mmwave} communications and the variability of data distributions, which remains an open issue.

In this paper, we introduce a dynamic layer-wise and clustering-based \ac{fl} algorithm to predict optimal beam sector selection in autonomous vehicles called \ac{sigla}.
The algorithm considers a layer sensitivity analysis to identify critical layers of the \ac{ml} model to reduce latency and communication overhead while decreasing communication failure probability during the wireless model transfer.
\ac{sigla} considers an intra-cluster algorithm to aggregate model layers within each cluster to prevent overfitting of ML models. 
\ac{sigla} also implements an inter-cluster algorithm to combine models from multiple clusters into a single model to represent a higher abstraction level, ensuring that the final model not only captures the specific characteristics of each cluster but also incorporates broader patterns observed across multiple clusters. We evaluate \ac{sigla} on a real-world multi-modal dataset collected from an autonomous car environment.

The proposed \ac{sigla} fundamentally advances prior FL approaches through three mmWave-specific innovations: (1) Dynamic layer sensitivity adaptation that prioritizes physical-layer parameters (beam angle, RSSI) over higher-level features, (2) Environment-aware clustering using real-time LIDAR spatial signatures to group vehicles by similar propagation conditions, and (3) Hybrid aggregation that combines intra-cluster specialization with inter-cluster generalization through attention-weighted layer fusion. 

The remainder of this article is organized as follows. \cref{sec:rw} details the current state-of-the-art. \cref{sec:system} describes the architecture, while \cref{sec:edafl} explains the \ac{sigla} algorithm. \cref{sec:exp_setup} introduces the experimental setup with its evaluation results.
Finally, \cref{sec:conclusions} concludes the paper.

\section{Related Works}
\label{sec:rw}

Salehi \etal{} \cite{salehi_flash_2022} proposed the FLASH algorithm to reduce sector selection time by leveraging local data processing, fusing multiple non-RF sensors as input to \ac{fl} in order to predict optimal communication sectors more efficiently than traditional methods.
FLASH-and-Prune extends FLASH by integrating model pruning within \ac{fl} to reduce communication overhead while improving scalability in \ac{mmwave} vehicular networks \cite{Salehi_Roy_Gu_Dick_Chowdhury_2024}. 
Traditional \ac{nn} pruning methods, such as MBP \cite{gale2019state}, eliminate weights with the smallest magnitudes.
Xue and Yang discussed the deployment of ultradense \ac{mmwave} networks, where \ac{fl} plays a critical role in managing beam alignments dynamically to cope with the fast-changing vehicular environment \cite{Xue_Yang_2023}.

Karimi \etal{} \cite{Karimi_Li_Li_2022} incorporated layer-wise adaptivity into local model updates using algorithms such as Fed-LAMB and Mime-LAMB, enhancing convergence speed and generalization performance across various datasets and model architectures for both \ac{iid} and non-\ac{iid} data distributions.
Lee \etal{} \cite{Lee_Zhang_Avestimehr_2023} proposed the FedLAMA scheme to adjust the aggregation interval on a layer-wise strategy based on the model discrepancy and communication cost, reducing the communication costs without significantly impacting model accuracy.
However, novel layer sensitivity analysis and layer selection approaches, together with clustering schemes, are required to improve the efficiency of FedLAMA in dynamic and non-\ac{iid} autonomous vehicle environments.

Briggs \etal{} \cite{Briggs_Fan_Andras_2020} enhanced \ac{fl} by introducing a \ac{hc}, which segments clients based on the similarity of their local updates to the global model.
Similarly, Ghosh \etal{} \cite{Ghosh_Chung_Yin_Ramchandran_2020} developed the iterative \ac{fl} clustering algorithm, which alternates between estimating user cluster identities and optimizing model parameters via gradient descent.

Unlike prior works that focus on either layer-wise aggregation \cite{Karimi_Li_Li_2022, Lee_Zhang_Avestimehr_2023} or clustering techniques \cite{Briggs_Fan_Andras_2020, Ghosh_Chung_Yin_Ramchandran_2020}, \ac{sigla} integrates dynamic clustering, layer-wise sensitivity analysis, and adaptive aggregation to reduce bandwidth requirements significantly. At the same time, it enhances the efficiency of the learning process. 
In this sense, \ac{sigla} extends the concept of \ac{nn} clustering by implementing a dynamic clustering mechanism that adapts to real-time changes in vehicular networks, improving the robustness and accuracy of sector beam selection. \ac{sigla} 's adaptive aggregation strategy tailors model updates based on the detected similarities and differences across vehicle clusters. Hence, \ac{sigla} offers a comprehensive and scalable solution to address the complexity of mmWave communications and the variability of data distributions, setting a new standard for FL in high-mobility environments.

\section{System Model and Preliminaries}
\label{sec:system}

\begin{figure}[!htb]
    \centering
    \includegraphics[width=0.48\textwidth]{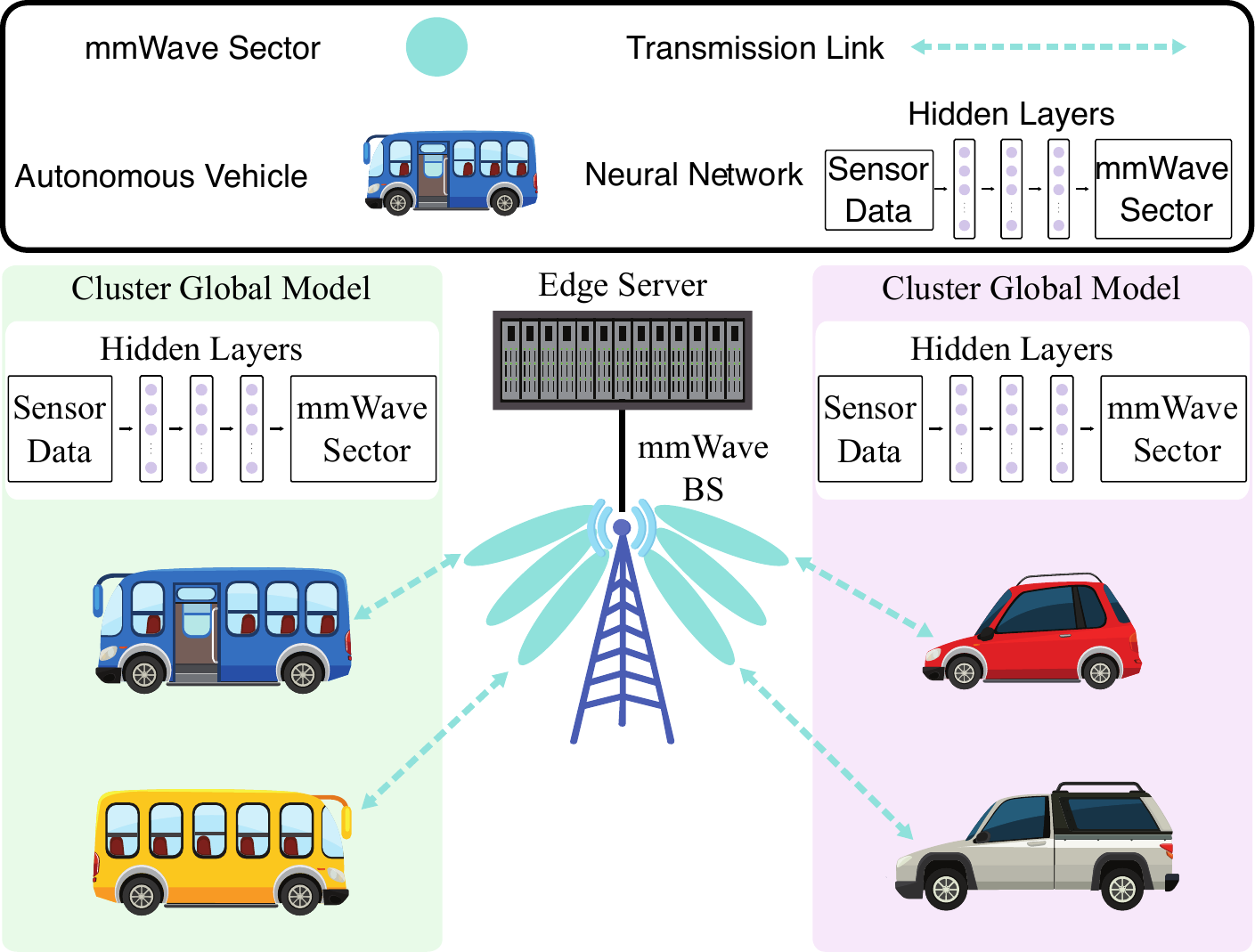}
    \caption{\ac{fl}-based Sector Selection Scenario}
    \vspace{-0.3cm}
    \label{fig:system}
\end{figure}

\cref{fig:system} illustrates the \ac{fl}-based sector selection over autonomous vehicle scenario. We assume the presence of \ac{mmwave} \ac{bs}, edge server, as well as a set of $N$ vehicles $n_i \in \{n_1, n_2, ..., n_N\}$ equipped with a set of sensors and \ac{mmwave} transceivers.
Upon the vehicle detecting a \ac{mmwave} \ac{bs}, it predicts the best \ac{mmwave} sector to connect to based on multi-modal sensor data, namely the vehicles' camera, LIDAR, and GPS sensors.
In this sense, each vehicle $n_i$ considers onboard units to train a \ac{nn} model and maintains its private local dataset $D_n$.
The datasets significantly vary in size, feature distribution, and label distribution, showing non-\ac{iid} characteristics.
Local datasets can be denoted as a 4-tuple of $D_n = \{X_{Camera}^n, X_{LIDAR}^n, X_{GPS}^n, X_{RF}^n \}^N_{n=1}$, where:
\begin{itemize}
\item $X_{Camera}^n \in \mathbb{R}^{S \times d^I_0 \times d^I_1}$
\item $X_{LIDAR}^n \in \mathbb{R}^{S \times d^L_0 \times d^L_1 \times d^L_2}$
\item $X_{GPS}^n \in \mathbb{R}^{S \times 2}$
\item $X_{\text{RF}}^n \in \mathbb{R}^{n\_sectors}$ is the \ac{RF} measurements capturing sector-specific signal quality metrics from the \ac{mmwave} transceiver.  
\end{itemize}

In this context, $\{d^I_0 \times d^I_1\}$ and $\{d^L_0 \times d^L_1 \times d^L_2\}$ represent the dimensions of the image matrix and the LIDAR point cloud, respectively. $X_{RF}$ denotes the signal strength measurements for all sectors detected by the vehicles' \ac{mmwave} interfaces.
GPS data can be characterized by a sequence of latitude and longitude samples.
Finally, $S$ indicates the number of samples in each local dataset $D_n$.
Each vehicle $n_i$ trains a local model $M_n$ with parameters $\theta_n$ on its dataset $D_n$, where
vehicles share their locally trained models $M_n$ and receive model updates from an edge server via reliable communication networks, including WiFi and 5G.

The \ac{ml} model $M$ consists of \( l \) layers denoted as \( M \equiv \{layer_1, layer_2, ..., layer_l\} \), where the weight matrix of its neurons characterizes each $layer_j$.
The system supports an \ac{fl} layer-wise approach to reduce the convergence time and improve accuracy \cite{Lee_Zhang_Avestimehr_2023}, which is an aggregation mode wherein vehicles transmit only a subset of layers $L_{selected}$  instead of transferring entire model's layers.
Each vehicle $n_i$ implements three different \ac{nn} models during the training phase. 
The first model relies on a \ac{cnn} to process the LIDAR point cloud data \(X_{LIDAR}^n\) and extracts spatial features such as distance and intensity of points.
The second model is a 2D \ac{cnn} for handling image data \(X_{Camera}^n\) and extracts features like edges, textures, and objects.
The third model implements a dense \ac{nn} for managing GPS data \(X_{GPS}^n\) to determine precise vehicle location coordinates.
Hence, the system uses different \ac{ml} models to handle each data type, simplifying the encoding of multi-modal data with different dimensionality characteristics.

\ac{fl} learns a parameter set $\theta$ to optimize sector selection for communication links between phased array antennas.
In this way, it maximizes communication efficiency while minimizing latency compared to IEEE 802.11ad and 5G-NR standards. 
After predicting the optimal sector, a vehicle shares it with the \ac{bs} via a control channel for sector selection. 
For instance, \ac{oran} allows the \ac{bs} to quickly use the predicted sector for \ac{mmwave} transmission. 
The optimal sector $t^*$ for a transmitter $tx$ configured at sector $s$ is computed based on Eq. \ref{fig:sector}, where $\theta$ is used to predict the sector quality metric $t_s$ for each candidate sector.

\begin{equation}
    t^* = \arg \max_{1 \leq n\_\text{sectors} \leq M} t_{s}(\theta)
    \label{fig:sector}
\end{equation}

\ac{fl} aims minimize the loss function \(L_n(\theta_n)\) the total number of samples across all vehicles $N$, as shown in Eq. \ref{fig:loss}.

\begin{equation}
    L(\theta) = \sum_{n=1}^N \frac{S}{I} L_n(\theta_n)
    \label{fig:loss}
\end{equation}

\section{\acf{sigla}}
\label{sec:edafl}

This chapter introduces the federated learning framework for \ac{mmwave} sector selection in autonomous vehicles, focusing on three key components:  
1) Layer sensitivity analysis (Subsection \ref{sec:layer_sensitivity}),  
2) Cluster-based model aggregation (Subsection \ref{sec:clustering}),  
3) Intra- and inter-cluster aggregation mechanisms (Subsections \ref{sec:aggregation_intra} and \ref{sec:aggregation_inter}).  

\subsection*{Overview}  
\begin{figure}[!htb]
    \centering
    \includegraphics[width=0.48\textwidth]{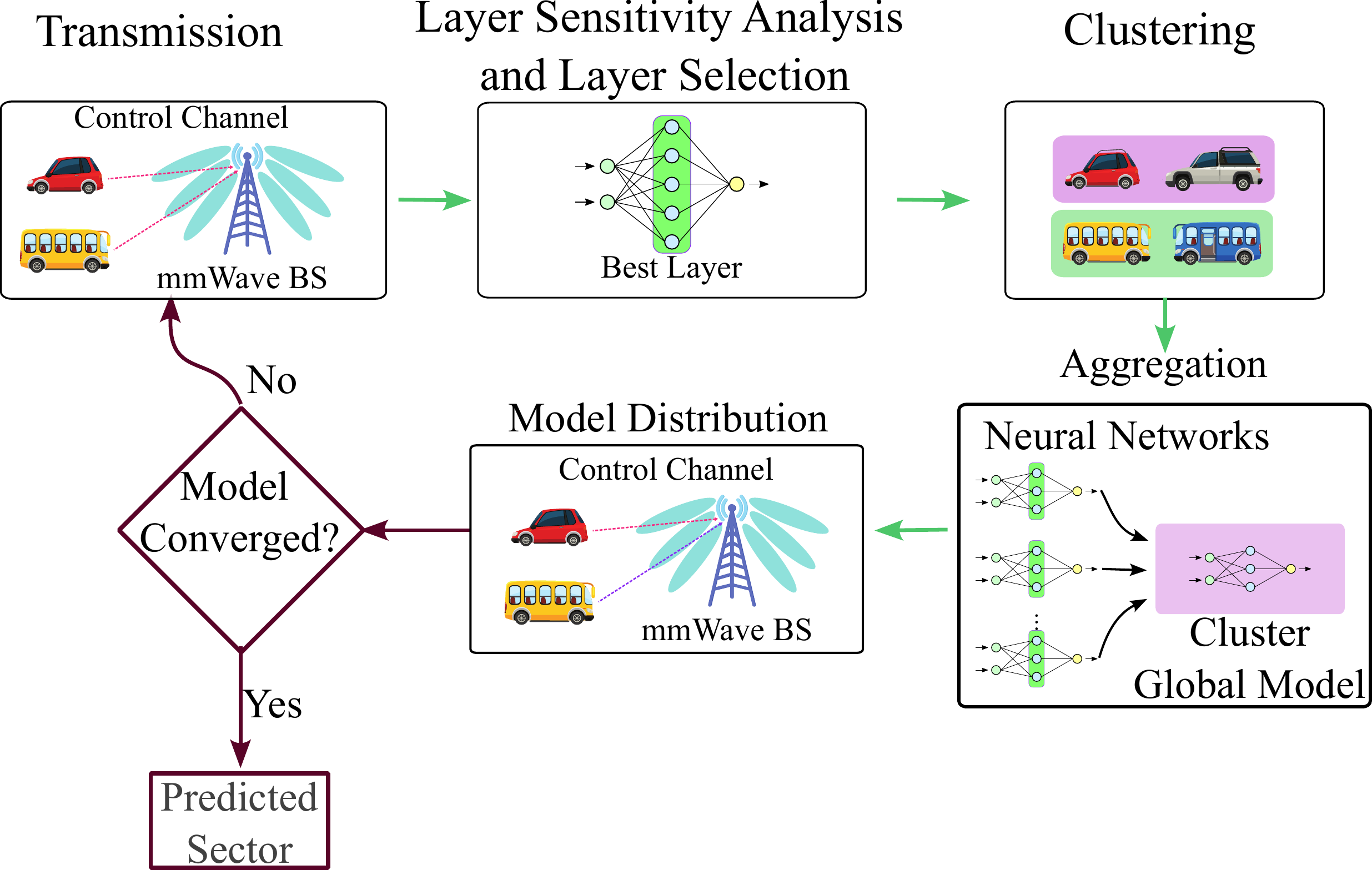}
    \caption{\ac{sigla} components and interactions}
    \label{fig:edafl}
\end{figure}

\Cref{fig:edafl} illustrates the components and interactions of the \ac{sigla} framework, which coordinates local training, model transfer, and aggregation to optimize \ac{mmwave} sector selection in dynamic autonomous vehicle environments. Initially, each vehicle \( n_i \) trains a local \ac{nn} model \( M_n \) using multi-modal sensor data (GPS, LIDAR, Camera) and transmits \( M_n \) to an edge server via the \ac{mmwave} \ac{bs}. The server employs cluster-based strategies to address participant churn: cluster inheritance handles new entrants, and contact time-weighted aggregation mitigates instability from transient vehicles.  

Subsequent steps involve layer sensitivity analysis to identify critical \ac{nn} layers (Subsection \ref{sec:layer_sensitivity}), clustering models with similar data distributions (Subsection \ref{sec:clustering}), and aggregating intra-/inter-cluster models (Subsections \ref{sec:aggregation_intra} and \ref{sec:aggregation_inter}). The refined global models are redistributed to vehicles via the \ac{mmwave} \ac{bs}, enabling iterative improvements until convergence. This cyclic process ensures adaptation to environmental dynamics, achieving accurate sector predictions for reliable \ac{mmwave} communication.  

\subsection{Layer Sensitivity Analysis and Layer Selection}
\label{sec:layer_sensitivity}

In contrast to other pruning and quantization methods, \ac{sigla} employs a layer-wise approach to determine the impact of a model layer on overall accuracy. In this sense, \ac{sigla} sends only important layers during each round, reducing the convergence time and improving accuracy.
We define the sensitivity of a \ac{nn} layer as the change in accuracy when a little noise is introduced to that layer.
The rationale behind layer sensitivity analysis lies in observing the performance degradation due to perturbations to a layer's parameters, determining the importance of such a layer to the model's accuracy, as discussed by Liu et al. \cite{Liu_Elkerdawy_Ray_Elhoushi_2021}.
In this sense, \ac{sigla} measures each layer's impact under controlled disturbance conditions, where the edge server adds zero-mean Gaussian noise \(\delta_{i,j} \sim \mathcal{N}(0, \epsilon^2 I)\) to the \(j^{th}\) layer's parameters of the \(i^{th}\) vehicle's model \(M_i\).
The parameter \(\epsilon\) is selected to balance the need for meaningful perturbation against the risk of excessively distorting the layer's functionality, while \(\theta_{i,1}, \dots, \theta_{i,j} + \delta_{i,j}, \dots, \theta_{i,l}\) denotes the weights of the model \(M_i\) with Gaussian noise added to the \(j^{th}\) layer's parameters.
Hence, \ac{sigla} identifies layers with a significant impact on accuracy, allowing the edge server to prioritize updates to the most important layers, enhancing the scalability and efficiency of the \ac{fl} process.
The importance score \(\lambda_j\) on the \(j^{th}\) layer of a given vehicle \(n_i\) is modeled based on Eq. \ref{eq:importance}.

\begin{equation}
    \small
    \lambda_j = \left|\text{Acc}(M_i) - \text{Acc}\left(M_i\left[\theta_{i,1}, \dots, \theta_{i,j} + \delta_{i,j}, \dots, \theta_{i,l}\right]\right)\right|
    \label{eq:importance}
\end{equation}

The edge server assesses the importance of each layer \(layer_j\) in the \ac{ml} model \(M_N\) through importance scores \(\lambda_j\).
A dynamic threshold \(\lambda_{\text{threshold}}\) is implemented upon verifying the current networking and computing resources at the network edges and is adjusted according to observed network conditions.
Hence, layers that meet or exceed the threshold are marked for transmission, and only important layers will be transmitted to the edge server.

The impact of layer sensitivity analysis and selection can be quantified by comparing the total number of \ac{nn} weights in all layers \(\Theta = \{\theta_1, \theta_2, \dots, \theta_J\}\) (where \(\theta_j\) represents the parameters of layer \(j\)) to the number of weights in selected layers \(\Theta_{\text{selected}}\). Here, \(|\theta_j|\) denotes the weight count of layer \(j\). The reduction factor measures how much the communication load is decreased through the layer-wise approach:

\begin{equation}
    \text{Reduction Factor} = \frac{\sum_{j \in L_{\text{selected}}} |\theta_j|}{\sum_{j=1}^{l} |\theta_j|}
    \label{fig:reduction}
\end{equation}

\subsection{Clustering}
\label{sec:clustering}

By grouping vehicles using \ac{cka} and \ac{hc} algorithms based on data distribution similarity, specialized models can be trained, leading to faster convergence and improved accuracy.
Specifically, \ac{cka} evaluates the functional similarity between model layers by measuring statistical dependence.
On the other hand, the \ac{hc} algorithm uses the similarity matrix generated by \ac{cka} to form clusters. The \ac{cka} implementation determines vehicle grouping based on model similarity post-training, as shown in Eq. \ref{eq:CKA}.
In its operation, vehicles send their model to the edge server, which clusters these parameters to group vehicles based on model similarity.

\begin{equation}
    \footnotesize
    \text{CKA}(X_{i,j}, X_{k,j}) = \frac{\text{HSIC}(X_{i,j}, X_{k,j})}{\sqrt{\text{HSIC}(X_{i,j}, X_{i,j}) \times \text{HSIC}(X_{k,j}, X_{k,j})}},
    \label{eq:CKA}
\end{equation}

where $X_{i,j}$ and $X_{k,j}$ represent the $j$-th model layer for vehicles $i$ and $k$.
\ac{HSIC} quantifies the dependence level between the models' layer activations, which is computed using kernel matrices $K$ and $L$, representing inner products of features transformed by a kernel function.
The edge server employs a polynomial kernel, defined by $K(x, y) = (x^\top y + c)^d$, where $c$ is a constant and $d$ is the kernel's degree.
The computed \ac{cka} values reflect functional similarities and are used to build a similarity matrix, providing insights into model pattern capture.

\ac{sigla} uses similarity matrix generated by \ac{cka} as input for the Dynamic Clustering Using Agglomerative \ac{hc} algorithm.
Initially, each vehicle is in a separate cluster, and the algorithm merges similar clusters based on similarity scores based on the following linkage methods.
\begin{inparaenum}[i)]
    \item Ward's method minimizes total within-cluster variance to achieve compact and spherical clusters.
    \item Complete Linkage helps when outliers are not a significant concern. 
    \item Average Linkage balances single and complete linkage by using average distances.
    \item Single Linkage is advantageous for large datasets to preserve the chaining effect in clusters.
\end{inparaenum}
In this way, we evaluate the silhouette scores and the Calinski-Harabasz index from each linkage method and choose the one with the best performance.
Precisely, the silhouette scores measure how similar each vehicle is to its cluster compared to others, as shown in Eq. \ref{eq:s}.

\begin{equation}
    s = \frac{b - a}{\max(a, b)},
    \label{eq:s}
\end{equation}

where \( a \) is the mean intra-cluster distance and \( b \) is the mean nearest-cluster distance. On the other hand, the Calinski-Harabasz index provides a criterion for determining the optimal number of clusters by maximizing the ratio between cluster variance and within-cluster variance. A higher value indicates better-defined clustering with compact, well-separated clusters, helping \ac{sigla} to select the optimal number of clusters for efficient vehicle grouping.

\subsection{Intra-Cluster Layer Aggregation}
\label{sec:aggregation_intra}

\ac{sigla} implements an Intra-Cluster Layer Aggregation algorithm to aggregate model layers within each cluster.
This algorithm considers the impact of layers to weight contributions from different vehicles, personalizing the aggregated model to cluster members while ensuring that it remains general enough to avoid overfitting.
Hence, by aggregating the layers within a given cluster, \ac{sigla} considers contributions from vehicles that do not belong to that cluster but have some similarity to its features, ensuring better model generalization.

After determining the number of clusters and participants, \ac{sigla} establishes the optimal number of iterations, denoted as $I_t$.
During aggregation, \ac{sigla} further considers the results when the clustering process is iterated $I_t + 1$ times.
The user labels and clusters obtained from this additional iteration are represented by $\iota_{+1}$ and $K_{+1}$, respectively.
By comparing the clusters formed at $I_t$ and $I_t + 1$, \ac{sigla} ensures a more robust and refined clustering outcome.

For each cluster $\kappa \in \mathcal{K}$, \ac{sigla} computes a weighted average of the local models of participating vehicles.
The weights are assigned based on the connectivity of the user layers within the cluster, i.e., similarity to other members within the cluster.
The averaging lets vehicles more representative of the cluster features influence the aggregated model, as shown in Eq. \ref{eq:theta}.

\begin{equation}
    \theta_{\kappa} = \frac{\sum_{i \in \kappa} w_i \theta_i}{\sum_{i \in \kappa} w_i},
    \label{eq:theta}
\end{equation}

where $w_i$ is a weight derived from the vehicle $i$'s similarity to other vehicles in cluster $\kappa$.

\subsection{Inter-Cluster Aggregation}
\label{sec:aggregation_inter}

\ac{sigla} implements an Inter-Cluster Aggregation algorithm to fine-tune the global model by combining related clusters into a single model representing a higher abstraction level.
For each primary cluster $\kappa \in \mathcal{K}$, its corresponding super-cluster $\kappa_{+1} \in K_{+1}$ is identified.
The models of the vehicles in $\kappa_{+1}$ are then aggregated using a similar weighted approach, ensuring broader cluster contributions as shown in Eq. \ref{eq:similar}.

\begin{equation}
    \theta_{\kappa_{+1}} = \frac{\sum_{j \in \kappa_{+1}} w_j' \theta_j}{\sum_{j \in \kappa_{+1}} w_j'},
    \label{eq:similar}
\end{equation}

where $w_j'$ reflects the relevance of each vehicle in $\kappa_{+1}$ to the vehicles in $K_{+1}$.
The final aggregation combines the models from $\kappa$ and $\kappa_{+1}$ to form the global model for cluster $C$ by averaging the parameters of $\kappa$ and $\kappa_{+1}$ as shown in Eq. \ref{eq:global}

\begin{equation}
    \Theta_{\kappa}^{\text{global}} = \frac{\theta_{\kappa} + \theta_{\kappa_{+1}}}{2}
    \label{eq:global}
\end{equation}

This robust and generalizable final model blends localized and extended patterns, enhancing predictive performance and reliability in dynamic vehicular environments.

\subsection{Algorithm Description}

The algorithm begins with each vehicle \( n_i \) initializing its local model \( M_n \) with parameters \( \theta_n \) (line \ref{alg:edafl:initialize}).
During each round \( t \) (line \ref{alg:edafl:rounds}), the edge server performs similarity measurement using \ac{cka} on selected layers (line \ref{alg:edafl:cka}).
Vehicles train their models locally to minimize loss \( L_n(\theta_n) \) (line \ref{alg:edafl:train}), and upload selected layers to the edge server (line \ref{alg:edafl:upload}).
The edge server conducts sensitivity analysis and applies HC to select and group important layers (lines \ref{alg:edafl:sensitivity}-\ref{alg:edafl:clustering}).
Within each cluster \( \kappa \), layers are aggregated (lines \ref{alg:edafl:aggregate-cluster1}-\ref{alg:edafl:aggregate-cluster2}), and these cluster models are further aggregated to update the global model \( \theta \), which is broadcasted back to vehicles (lines \ref{alg:edafl:global-aggregate}-\ref{alg:edafl:broadcast}).
Each vehicle updates its local model with global parameters (line \ref{alg:edafl:update-local}).
For sector selection, each vehicle predicts optimal sector \( t^* \) using the updated model and communicates it to the BS (lines \ref{alg:edafl:predict-sector}-\ref{alg:edafl:communicate-sector}).

\section{Performance Evaluation}
\label{sec:exp_setup}

\subsection{Simulation Environment}
\label{sec:simulation_env}

We conducted simulations using TensorFlow 
and Keras 
on a server with 13th Gen Intel i9-13900K, 128GB RAM, and two NVIDIA GeForce RTX 4090 GPUs. We used the FLASH dataset \cite{salehi_flash_2022}, which includes data from a 2017 Lincoln MKZ Hybrid vehicle equipped with GPS, a GoPro HERO4 camera, and a Velodyne VLP-16 LIDAR sensor. In the dataset, vehicles traveled along a two-way paved alley flanked by tall buildings in Boston City. Two TP-Link Talon AD7200 routers are positioned at the roadside base station and on the vehicle, operating at 60 GHz, provided RF ground truth including \ac{rss} at the receiver.

Parameter selection balanced mmWave vehicular constraints and federated learning requirements, with 0.002 learning rate via grid search (0.0001-0.01); 32 batch size from GPU memory constraints and LIDAR data dimensions; noise empirically set as 10\% of layer weights.

\begin{algorithm}[!htb]
\label{alg:edafl}
\caption{\ac{sigla} for beam sector selection}
\small
\SetAlgoLined
\KwData{Each vehicle $n_i$ has a local dataset $D_n$.}
\KwResult{Optimal \ac{mmwave} sector.}
{Initialization:}\\
\For{each vehicle $n_i$}{
Initialize local model $M_n$ with parameters $\theta_n$\; {\label{alg:edafl:initialize}}
}
\BlankLine
{Local Training Loop:}\\
\For{each round $t = 1, 2, \dots, T$}{ {\label{alg:edafl:rounds}}
Edge server performs clustering\; {\label{alg:edafl:cka}}
\For{each vehicle $n_i$}{
Train $M_n$ on $D_n$ to minimize $L_n(\theta_n)$\; {\label{alg:edafl:train}}
Upload $L_{selected}$ to the edge server\; {\label{alg:edafl:upload}}
}
\BlankLine
{Layer Selection and Clustering at Edge Server:}\\
Perform sensitivity analysis\; {\label{alg:edafl:sensitivity}}
Apply  HC based on similarity scores\; {\label{alg:edafl:clustering}}
\BlankLine
{Aggregation within Clusters:}\\
\For{each cluster $\kappa$}{
Aggregate selected layers within cluster: $\theta_\kappa \leftarrow \sum_{i \in \kappa} \frac{w_i}{\sum_{j \in \kappa} w_j} \theta_i$\; {\label{alg:edafl:aggregate-cluster1}}
} {\label{alg:edafl:aggregate-cluster2}}
\BlankLine
{Global Aggregation and Broadcast:}\\
Aggregate cluster models to update global model $\theta$\; {\label{alg:edafl:global-aggregate}}
Send updated global model $\theta$ back to all vehicles\; {\label{alg:edafl:broadcast}}
\For{each vehicle $n$}{
Update local model with global parameters: $M_n \leftarrow \theta$\; {\label{alg:edafl:update-local}}
}
}
\BlankLine
{Sector Selection:}\\
\For{each vehicle $n_i$}{
Predict the optimal sector $t^*$ using updated $M_n$\; {\label{alg:edafl:predict-sector}}
Transmit the selected sector $t^*$ to the \ac{bs} via control channel\; {\label{alg:edafl:communicate-sector}}
}
\end{algorithm}

The dataset has synchronized multi-modal data divided into four main categories and 21 scenarios (LOS and three NLOS conditions). The \ac{nlos} scenarios consists of pedestrians, static vehicles, and moving vehicles serving as obstacles for the signal. Each scenario comprises ten episodes, effectively representing data from 10 vehicles, each with 21 unique scenarios as their local dataset.
The four main categories are defined as follows \cite{salehi_flash_2022}:
LOS passing (\ie, Cat 1 in the plot): Vehicle passes through clear LOS.
NLOS pedestrian (\ie, Cat 2): Pedestrian obstructs LOS with variations in movement.
NLOS static car (\ie, Cat 3): Static car obstructs LOS in various positions.
NLOS moving car (\ie, Cat 4): Moving car crosses LOS with different speeds and lane positions.

Each scenario contains ten trials, representing data from 10 vehicles, divided into 80\% training, 10\% validation, and 10\% test sets. The global test dataset combines the remaining 10\% of each vehicle's local data, totaling 25,456 training samples, 3,180 validation samples, and 3,287 global test samples.

Each vehicle relies on a multi-modal \ac{nn} model with three submodels for image, lidar, and GPS data:
\begin{inparaenum}[i)]
    \item Image submodel consists of two convolutional layers with max-pooling and batch normalization, followed by dense layers with dropout.
    \item LIDAR submodel considers 3D convolutional layers with max-pooling and batch normalization, followed by dense layers with dropout.
    \item GPS submodel has dense and dropout layers.
\end{inparaenum}

We evaluated beam sector selection protocols including Centralized Learning, FLASH \cite{salehi_flash_2022}, FLASH-and-Prune \cite{Salehi_Roy_Gu_Dick_Chowdhury_2024}, MBP \cite{gale2019state}, FedLAMA \cite{Lee_Zhang_Avestimehr_2023}, and \ac{sigla}. We also compared prediction time for traditional \ac{mmwave} beam selection using IEEE 802.11ad with our approach.
We considered the following evaluation metrics:
{Accuracy} as the percentage of correct classifications,
{convergence time} as the time to reach a plateau in accuracy (in epochs),
{number of parameters transmitted} reflecting communication efficiency,
{number of models sent and received} reflecting communication strategy robustness,
{rate of successful model transmissions} means the percentage of successful data transmissions across training rounds.
We consider the successful model transmissions as the probability of transmitting a deep learning model with 29,833,376 parameters (approx. 113.81 MB) over an IEEE 802.11ad network \cite{Chandra_2017, n2012wigigieee80211ad, 8839948}.

\begin{figure*}
    \centering
    \begin{subfigure}[b]{0.36\textwidth}
        \includegraphics[width=\textwidth]{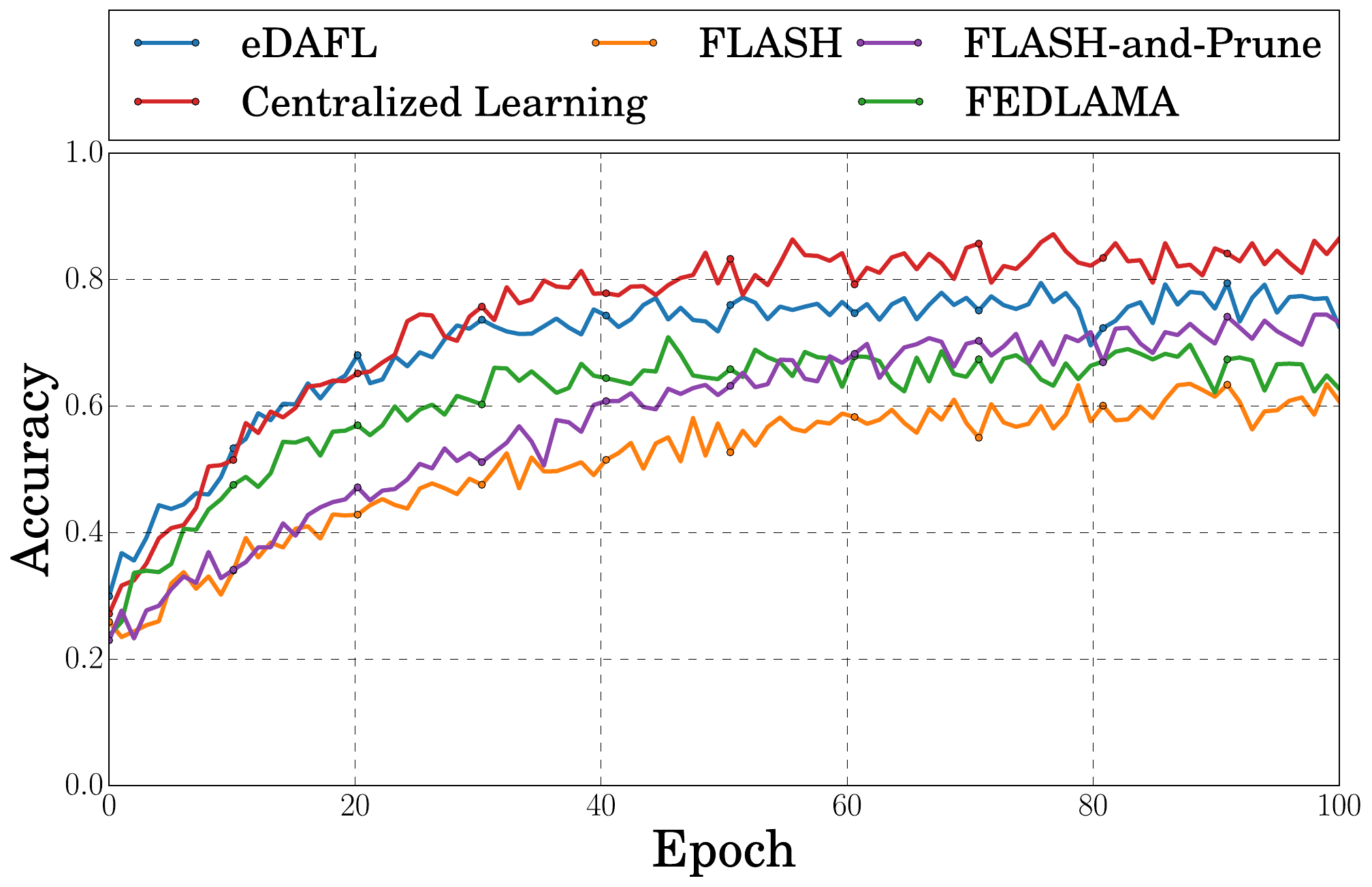}
        \caption{Acuracy over epochs}
        \label{fig:convergence}
    \end{subfigure}
    \begin{subfigure}[b]{0.42\textwidth}
        \includegraphics[width=\textwidth]{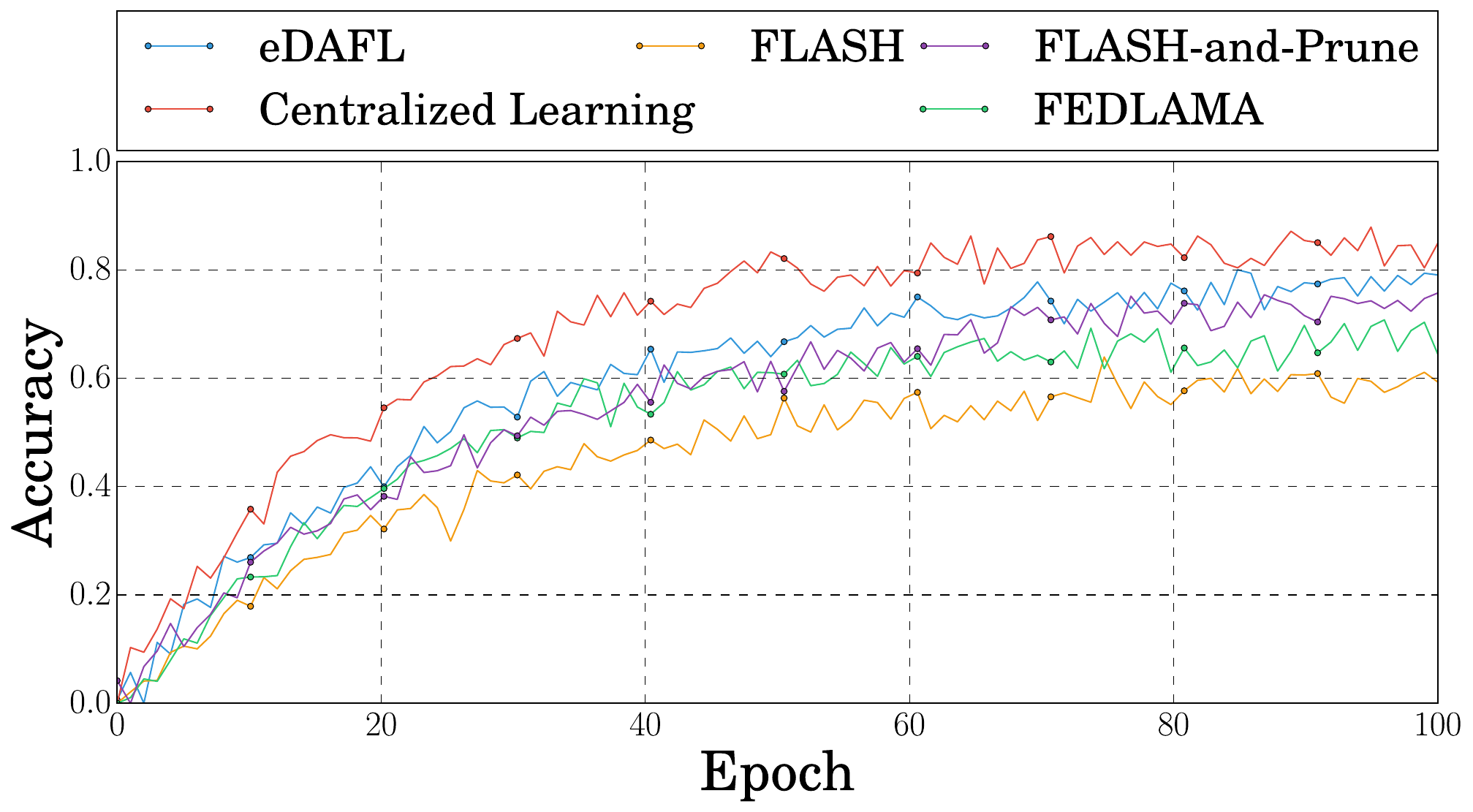}
        \caption{Top-1 Convergence}
        \label{fig:top1-convergence}
    \end{subfigure}
    \begin{subfigure}[b]{0.77\textwidth}
        \includegraphics[width=\textwidth]{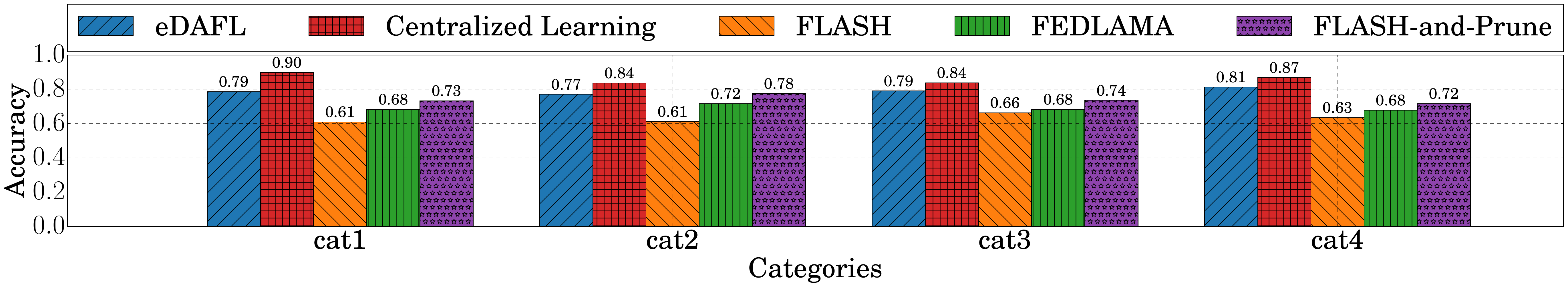}
        \caption{Accuracy comparison across different data categories}
        \label{fig:accuracy-category}
    \end{subfigure}
    \caption{Accuracy Results for the tested algorithms}
    \label{fig:acc_curves}
\end{figure*}

\subsection{Evaluation Results}
\label{sec:results}

\Cref{fig:convergence} shows the evolution of the prediction accuracy on the test dataset for all evaluated sector selection protocols.
We observe that centralized learning shows the fastest convergence due to the availability of the entire dataset in a centralized location.
However, centralized learning leads to high latency and communication costs for transferring the user data and poses privacy concerns as sensitive data could be intercepted.
On the other hand, \ac{sigla} performs better than the tested FL algorithms to predict and select the best mmWave sector for a vehicle to connect to, where \ac{sigla} provides results closer to centralized learning.
For instance, \ac{sigla} has a final accuracy of 8.14\% lower than Centralized Learning but higher by 25.40\%, 14.49\%, and 6.76\% compared to FLASH, FEDLAMA, and FLASH-and-Prune, respectively. \ac{sigla} 's higher final accuracy can be attributed to its clustering mechanism, with intra-clustering for handling non-\ac{iid} data distributions and inter-clustering for better model generalization.
Such clustering modules are absent in the other compared algorithms, and thus, they may struggle with non-IID data distributions.

\Cref{fig:top1-convergence} presents the Top-1 accuracy (\ie, highest accuracy to predict the mmWave sector) for the evaluated protocols over 100 rounds. 
We conclude that \ac{sigla} demonstrates a steady improvement, with final accuracy 6.86\% lower than centralized learning, while it is 4.42\% higher than FLASH-and-Prune.
FEDLAMA, although slower to converge, reaches an accuracy of around 0.64, 18.9\% lower than \ac{sigla}.
Despite its quick initial rise, FLASH stabilizes at a lower accuracy, indicating a faster but less accurate learning process.
FLASH-and-Prune achieves a final accuracy lower by 3.7\%.
\Cref{fig:accuracy-category} illustrates the accuracy of the tested protocols across the four different data categories. We can observe that \ac{sigla} shows consistent performance across categories, closely followed by FLASH-and-Prune and FedLAMA, which generally perform well.
FLASH exhibits the lowest accuracy across all categories, indicating faster convergence.
While FLASH-and-Prune achieves higher accuracy than FLASH, it is inferior to \ac{sigla} across all data categories.
Hence, \ac{sigla} improved the performance of each data category, even considering their different scenarios and features
since it considers an inter-cluster and intra-cluster aggregation to provide personalized models with better generalization.

\begin{figure*}
    \centering
    \begin{subfigure}[b]{0.3\textwidth}
        \includegraphics[width=\textwidth]{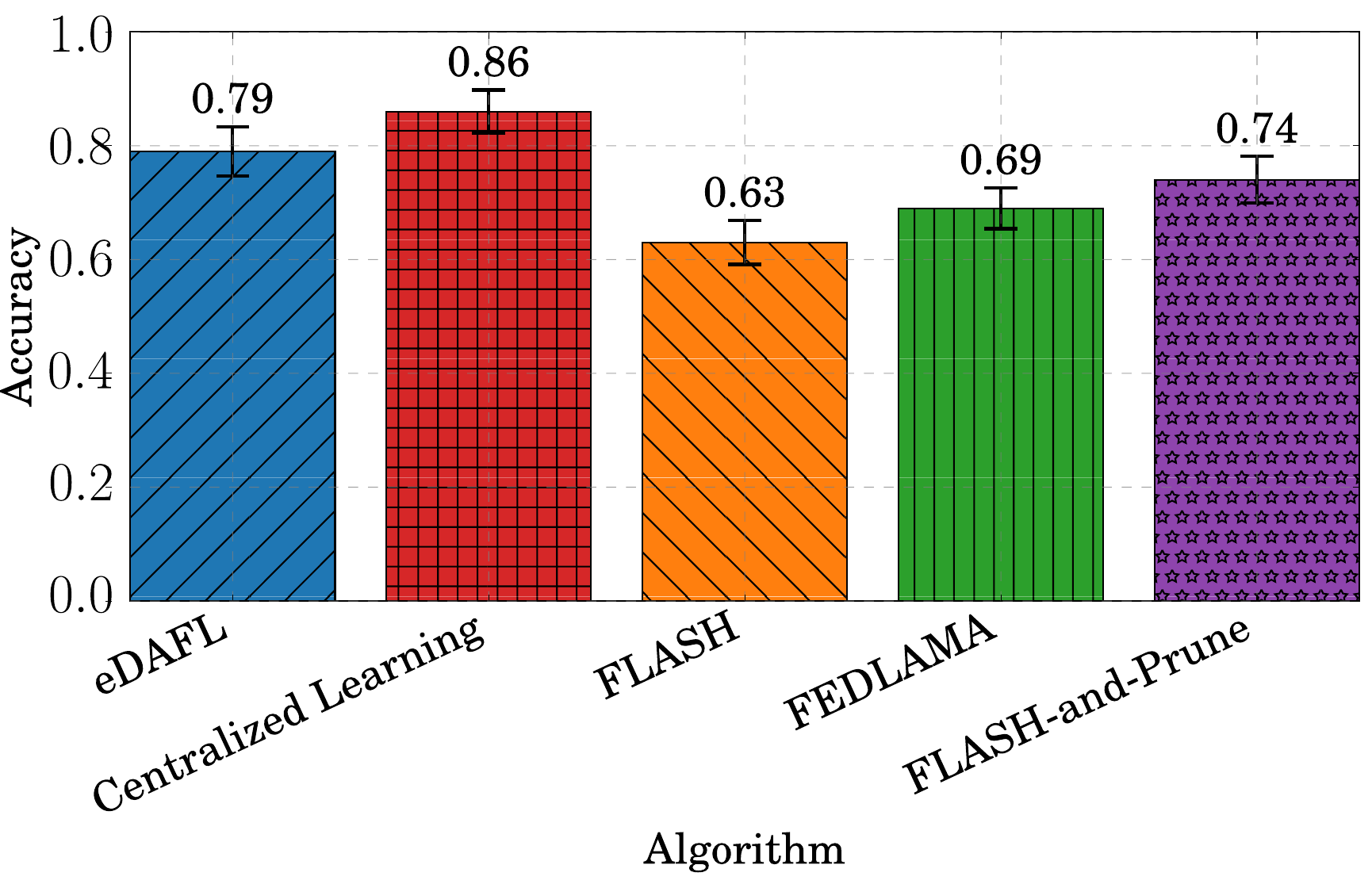}
        \caption{Final Accuracy}
        \label{fig:final-accuracy}
    \end{subfigure}
    \begin{subfigure}[b]{0.28\textwidth}
        \includegraphics[width=\textwidth]{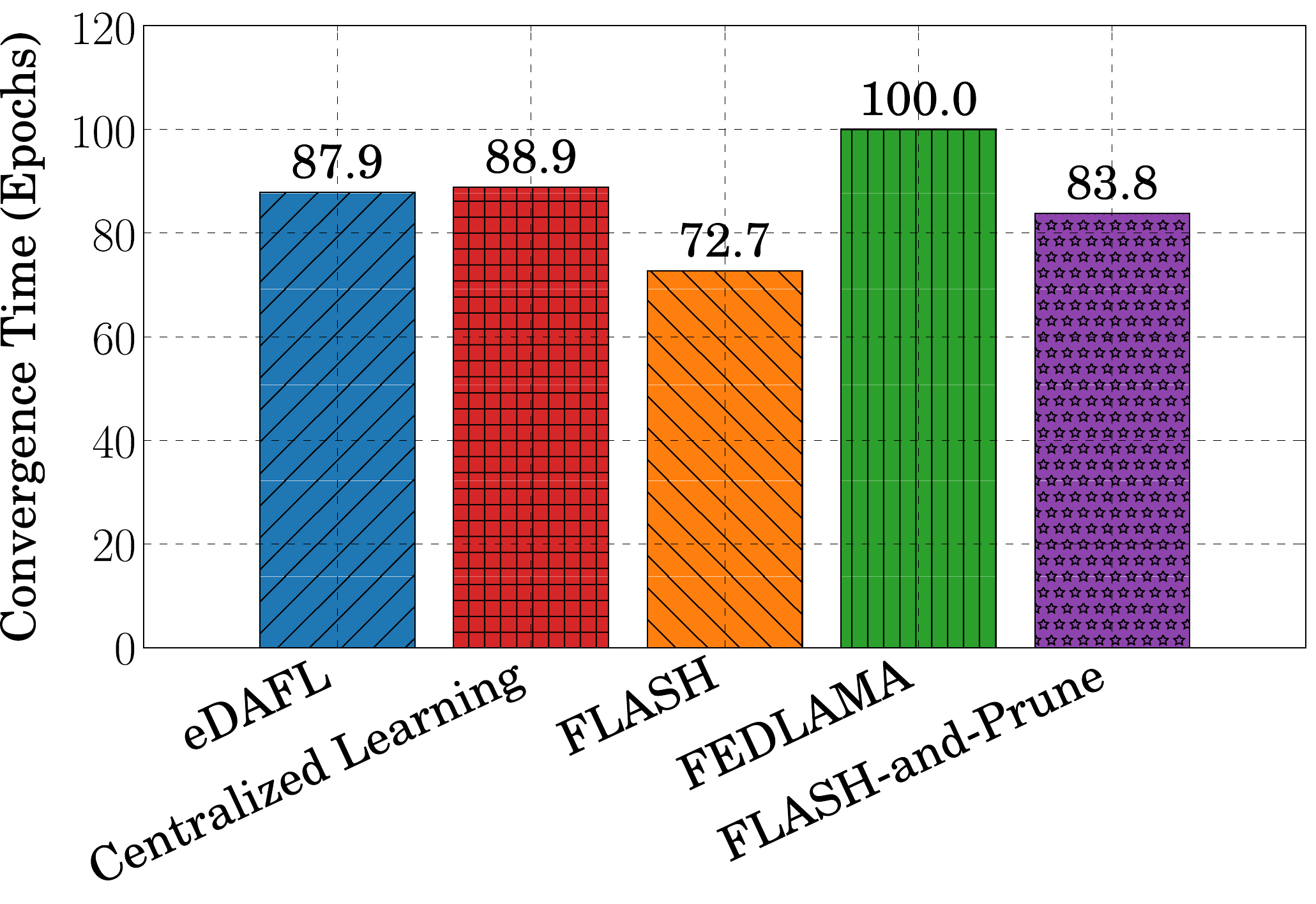}
        \caption{Convergence time}
        \label{fig:convergence-time}
    \end{subfigure}
    \begin{subfigure}[b]{0.3\textwidth}
        \includegraphics[width=\textwidth]{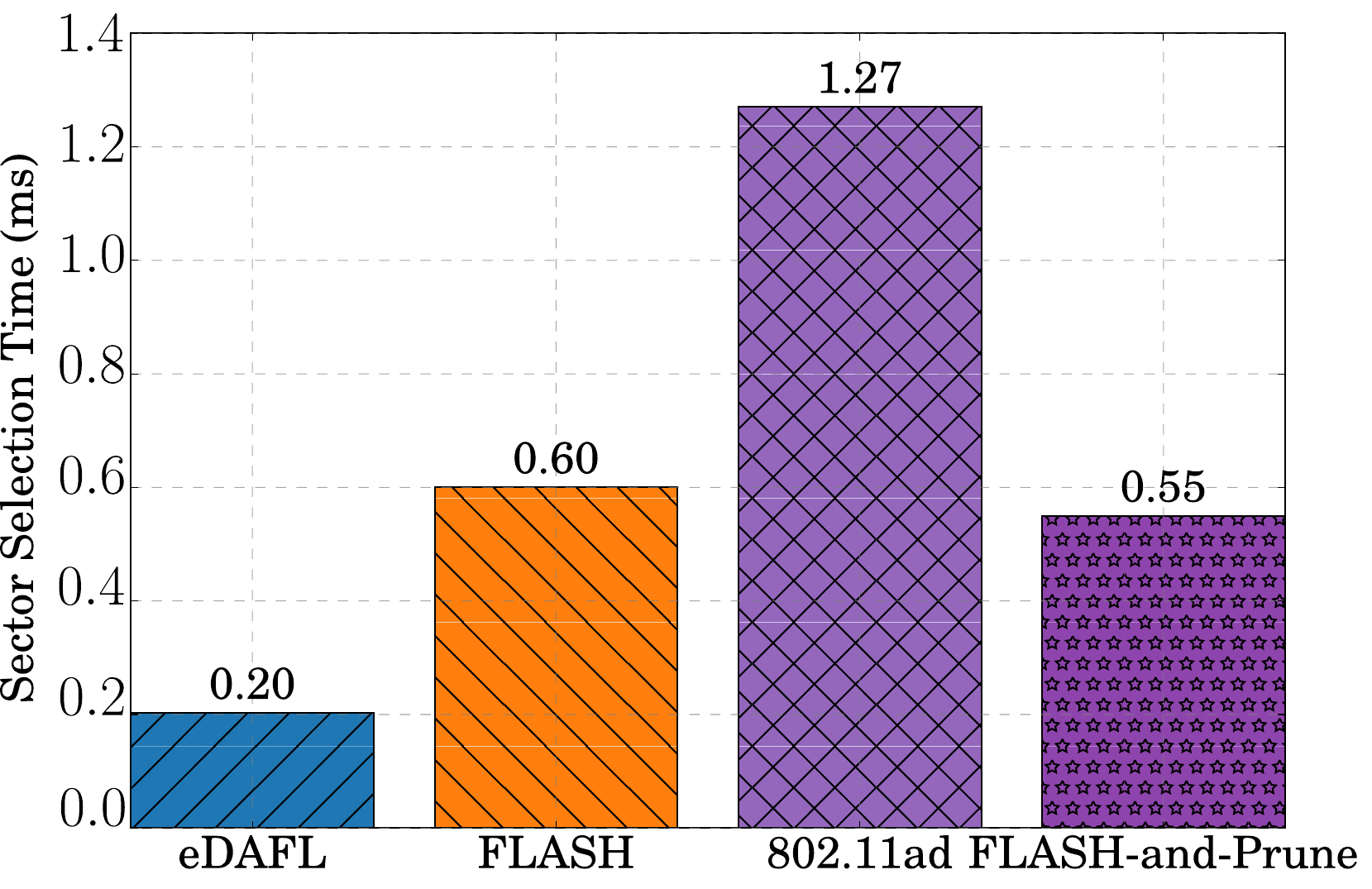}
        \caption{Sector Selection Latency}
        \label{fig:prediction-time}
    \end{subfigure}
    \caption{Final accuracy, Convergence time, and Sector Selection Latency for the evaluated algorithms}
    \label{fig:acc_curves}
\end{figure*}

\Cref{fig:final-accuracy} depicts the final accuracy for the analyzed algorithms.
This result highlights the performance gap between centralized and \ac{fl} approaches, where \ac{sigla} is the most effective \ac{fl} algorithm.
\Cref{fig:convergence-time} shows the convergence times for the analyzed algorithms, where 
FedLAMA takes the longest time.
These results illustrate the trade-offs between convergence speed and final accuracy, with FLASH being the fastest but least accurate. \ac{sigla} and Centralized Learning balance convergence times and higher accuracy.
The effective convergence of \ac{sigla} is due to its dynamic clustering and adaptive layer selection, ensuring efficient learning.
\Cref{fig:prediction-time} illustrates the inference time performance of \ac{sigla}, FLASH, and IEEE 802.11ad.
This evaluation asserts the viability of an ML-based sector selection protocol compared to the traditional sector search of IEEE 802.11ad. \ac{sigla} achieves the lowest inference time at 0.20 ms, outperforming FLASH (0.60 ms) and IEEE 802.11ad (1.27 ms). 
The poor performance of IEEE 802.11ad is due to its exhaustive sector search method, which involves bi-directional packet transmissions to investigate every possible sector, leading to significant delay by avoiding extensive searches. \ac{sigla} and FLASH use \ac{nn}s to predict the optimal sector, reducing the selection time. However, \ac{sigla} can shorten the sector selection time, improving communication quality for dynamic and mobile autonomous vehicles.

Reducing communication overhead (downlink and uplink interfaces) for model sharing is critical in dynamic autonomous vehicle environments. Using the \textit{float16} data type, we find that \ac{sigla} incurs 8.78MB (uplink) and 7.92MB (downlink) overhead per iteration, compared to 9.43MB/8.54MB for FLASH-and-Prune and 9.34MB/8.58MB for FedLAMA. \ac{sigla} converges faster due to its layer-wise clustering-based adaptive scheme, efficiently reducing data transmission while enhancing model accuracy.

\begin{figure*}
    \centering
    \begin{subfigure}[b]{0.32\textwidth}
        \includegraphics[width=\textwidth]{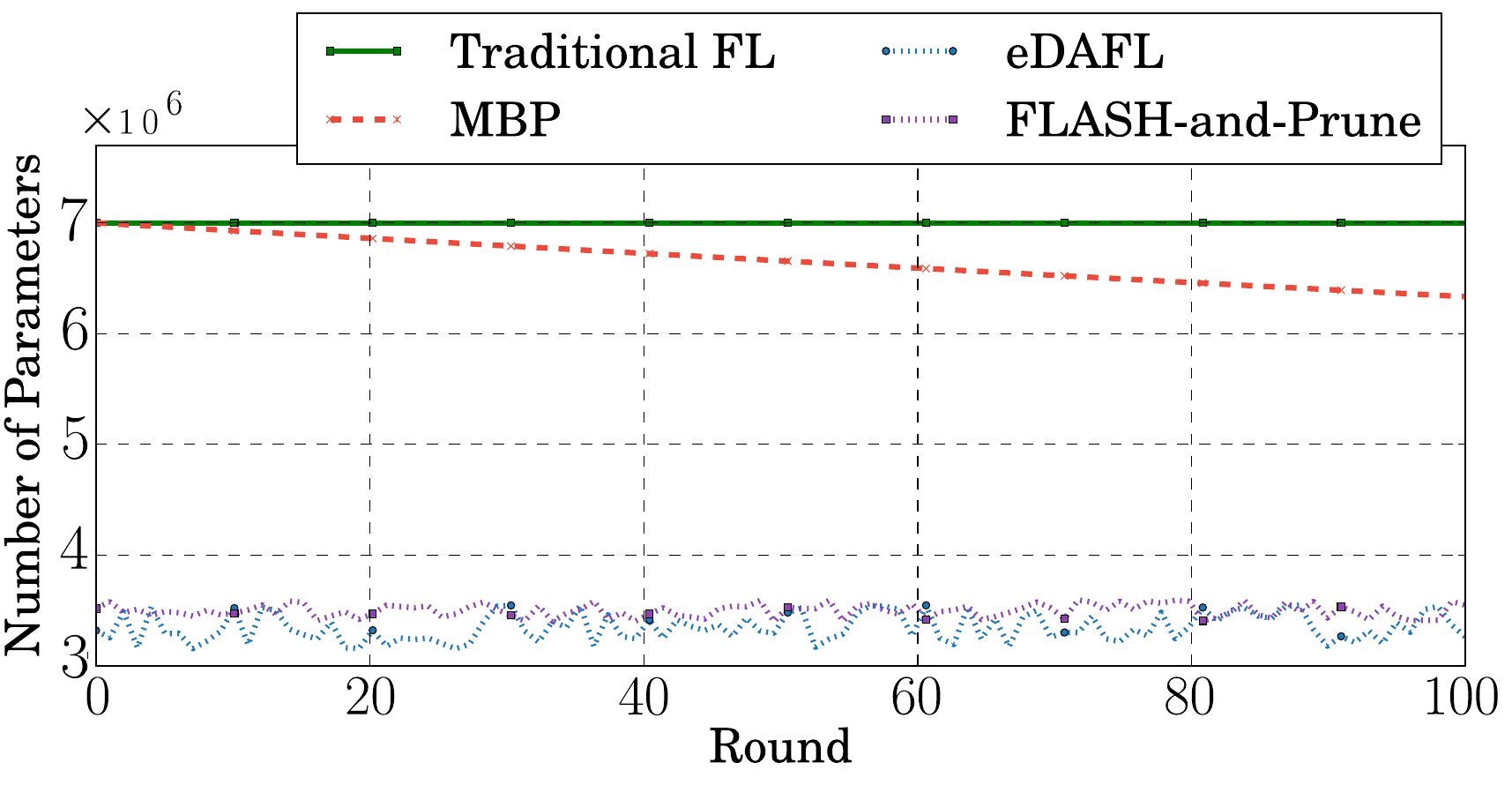}
        \caption{Number of Parameters transmitted}
        \label{fig:parameters}
    \end{subfigure}
    \begin{subfigure}[b]{0.3\textwidth}
        \includegraphics[width=\textwidth]{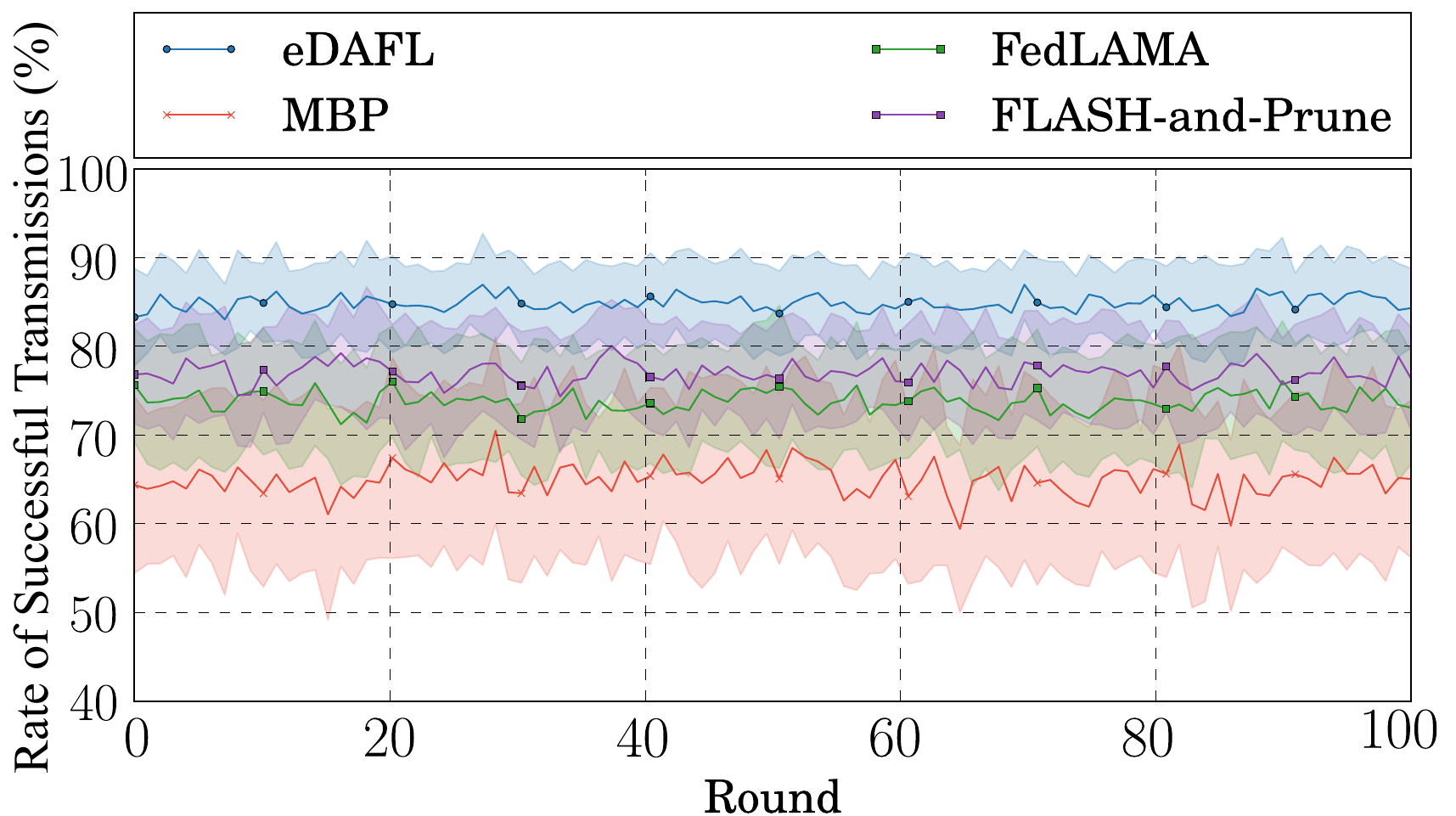}
        \caption{Rate of successful model transmissions}
        \label{fig:successful-transmissions}
    \end{subfigure}
    \begin{subfigure}[b]{0.3\textwidth}
        \includegraphics[width=\textwidth]{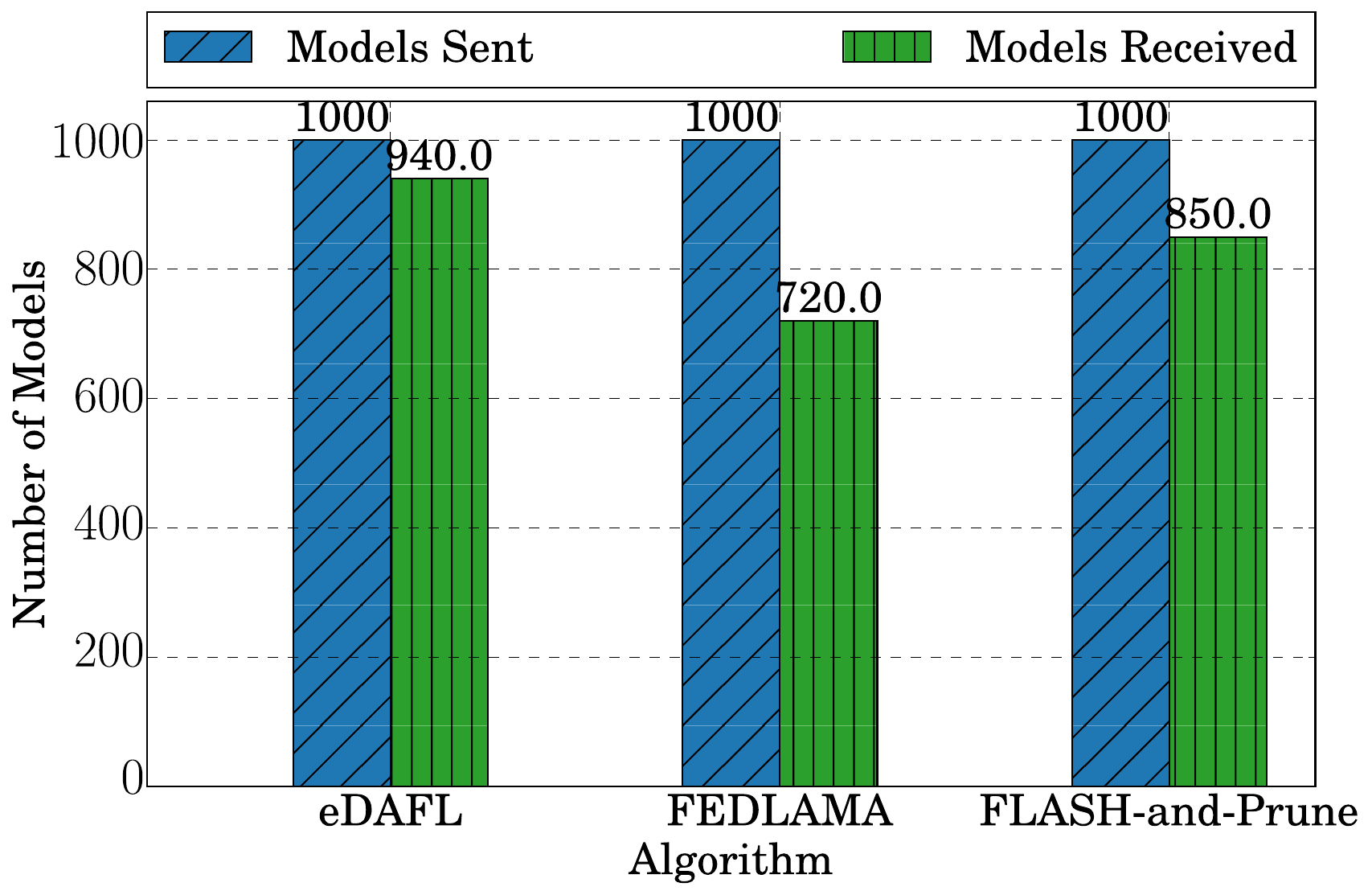}
        \caption{Number of Models Sent and Received}
        \label{fig:models-sent-received}
    \end{subfigure}
    \caption{Evaluation Results for the tested algorithms}
    \label{fig:acc_curves}
\end{figure*}

\Cref{fig:parameters} compares the number of parameters transmitted per aggregation round in traditional FL, MBP, and \ac{sigla}.
FedLAMA and Centralized Learning are not evaluated since they do not reduce the number of parameters sent.
Traditional \ac{fl} consistently transmits a more significant number of parameters.
MBP attempts to filter out weights.
While MBP decreases the number of \ac{ml} model weights sent, it lags behind FLASH-and-Prune and \ac{sigla}. \ac{sigla} reduces the number of transmitted parameters, showing a better reduction and enhancing communication efficiency. \ac{sigla} transmits 52.20\% fewer parameters than traditional FL, and 4.36\% fewer than FLASH-and-Prune.
Similarly, FLASH-and-Prune can also reduce the number of parameters transmitted to similar performance, but it lags behind \ac{sigla} in terms of prediction accuracy.
This implies that \ac{sigla} 's selection of the most important model layers and \ac{iid} clustering achieve similar parameter reduction to FLASH-and-Prune while not compromising prediction accuracy.
\ac{sigla} 's efficient layer selection and clustering techniques minimize the number of parameters transmitted, reducing communication costs while maintaining high accuracy.

\Cref{fig:successful-transmissions} shows the success rate of model transmissions during the \ac{fl} process. \ac{sigla} achieves a 94\% success rate, compared to 85\% for FLASH-and-Prune and 72\% for FedLAMA. \ac{sigla} prioritizes transmitting critical model layers within available contact time and bandwidth, ensuring higher success rates. In contrast, FedLAMA implements an iterative approach and transmits layers individually, reducing overhead but causing delays and potential losses. FLASH-and-Prune and FedLAMA also reduce model size effectively, but \ac{sigla} uses network resources more efficiently for model transmissions.

\Cref{fig:models-sent-received} compares the number of models sent and received by \ac{sigla} and FedLAMA. \ac{sigla} achieves 940 successful transmissions out of 1000 (\(R_s = 94\%\)) by prioritizing critical layers within the available contact time, outperforming FedLAMA's 720 transmissions (\(R_s = 72\%\)) and FLASH-and-Prune. Using a layer-wise transmission mechanism, \ac{sigla} reduces data transfer requirements and increases the probability of correctly receiving the model. Its sensitivity analysis identifies essential ML model layers for transfer, reducing overhead and failure rates in wireless model transmission.

\section{Conclusions}
\label{sec:conclusions}

We introduced the enhanced Dynamic Adaptive Federated Learning (\ac{sigla}) algorithm for beam sector selection in autonomous vehicle networks.
\ac{sigla} reduces network overhead and latency while enhancing the accuracy of beam sector selection.
Clustering using the \ac{cka} similarity metric and \ac{hc} optimizes intra-cluster aggregation, while lower-weight inter-cluster contributions enhance system efficiency with non-IID datasets.
Our contributions include an adaptive hierarchical clustering mechanism, a layer sensitivity analysis technique, and a soft clustering mechanism.
Together, these innovations enhance model accuracy and reduce network overhead.
Performance evaluation results demonstrate that \ac{sigla} improves model accuracy by approximately 6.76\% compared to FLASH-and-Prune and reduces inference time by 84.04\% compared to IEEE 802.11ad, with model size reduction of up to 52.20\% compared to traditional methods. 

\bibliographystyle{ieeetr}
\bibliography{bib}
\end{document}